\documentclass[conference]{IEEEtran}
\ifCLASSINFOpdf
\else
\fi
\usepackage{url}
\usepackage[]{fancyhdr} %
\newcommand{\changefont}{\fontsize{7}{8}\selectfont}
\usepackage{amsmath}
\usepackage{graphicx}
\usepackage{placeins}
\usepackage{xcolor}

\IEEEoverridecommandlockouts
\begin{document}
\title{Using Active Distribution Network Flexibility to Increase Transmission System Voltage Stability Margins}
\author{\IEEEauthorblockN{Giorgos Prionistis, Costas Vournas}
\IEEEauthorblockA{National Technical University of Athens\\Athens, Greece\\
gprionistis@power.ece.ntua.gr}}

\maketitle
\thispagestyle{fancy}
\pagestyle{fancy}

\begin{abstract}
The increasing penetration of Distributed Energy Resources (DER) in the distribution network creates new challenges in the operation of both the transmission and the distribution network. However, the controllability of the converter interfaced devices (CIG), also unveils opportunities for flexible operation and provision of ancillary services with or without economic incentives. The main scope of this work is to create a framework in order to calculate the operational flexibility of an Active Distribution Network and use it to address a centralized Optimal Power Flow Problem by the Transmission System Operator, and in particular the Voltage Stability Margin maximization. Two different approaches are proposed to calculate the Flexibility Region (FR) in the PQ plane, and the centralized optimization is applied to simple and more complex transmission test systems and feeder configurations.
\end{abstract}


\IEEEpeerreviewmaketitle

\section{Introduction}
The Distribution Network and it's operation is changing drastically, since more Converter Interfaced Generation (CIG), also known as Inverter Based Generation (IBG) and more generally Inverter Based Resources (IBR), such as Battery Energy Storage Systems (BESS) and flexible loads are connected turning them into networks with active and controllable components. This gives rise to the notion of Active Distribution Network (ADN). ADNs introduce more complexity but also new opportunities for support and ancillary services from the lower voltage side to the transmission grid. If correct communication is established between the Distribution System Operator (DSO) and the Transmission System Operator (TSO), even traditional transmission system problems such as frequency and voltage control, or congestion management can be tackled by exploiting the distribution network resources, without disrespecting the secure operational limits of the distribution network~\cite{hatziargyriou2017contribution}. 

Distribution feeders can be long and complex, consisting of Medium (MV), as well as Low (LV) Voltage levels. As a result, hundreds of feeders would create a very demanding computational problem, if modelled and simulated together with the transmission system. Moreover, most TSOs don't have access to the distribution network data. As a result, distribution feeders are usually treated from the TSOs as equivalent active and reactive power injections at the High Voltage (HV) side of HV/MV substations. 
As already stated above, a minimum degree of communication and coordination has to be established in order to request support from the ADNs~\cite{GIVISIEZ2020106659}. The Swiss TSO for example, already provides incentives for reactive power provision from the distribution side under certain circumstances~\cite{7038467}. Minimum requirements are also stated in VDE-AR-N 4141-1~\cite{VDE} for the TSO/DSO coordination.

The term operational flexibility or flexibility region of ADNs has been introduced in the recent literature~\cite{LI2020105872,8442865,8291006,9495085} to describe the amount of the regulating capacity that ADNs can provide in terms of active and reactive power, while satisfying their own operational constraints. Operational flexibility is translated into the active and reactive power exchange in the point of common coupling (PCC) between the transmission and the distribution network and the ability of each ADN to control these flows through it's own resources. Alternatively the same service is referred to as Reserve~\cite{8917723}, or Ancillary Service~\cite{CAPITANESCU2018226} Provision. 

Various methods and formulations have been proposed to identify and calculate ADN flexibility. More specifically, in~\cite{LI2020105872}, authors try to define and quantify this operational flexibility provision and availability of ADNs using the linearized DistFlow model and Monte Carlo simulations. The time-dependent flexibility in order to control the TSO-DSO interconnection power flows is calculated in ~\cite{8442865}, by determining a feasible operating region with a random number sampling approach. Authors of ~\cite{8917723}, propose a method to define a convex reserve provision capability area of an ADN by creating a linear scenario-based optimization problem and searching in multiple directions, starting from a base operating point. The works in~\cite{8291006},~\cite{CAPITANESCU2018226}  also define the flexibility area in the PQ plane, but using non-linear OPF formulations, with the former one also including different levels of flexibility costs. In the aforementioned works, authors ensure implicitly or explicitly the convexity of the flexibility region but that is not always the case. In ~\cite{9495085}, the authors address this issue and define a convex core of the non-convex flexibility region. 
Finally, in \cite{8810638} a distinction is made between the terms "flexibility" and "feasibility", while in~\cite{9485097} different flexibility regions are defined depending upon the services provided.

This paper focuses on the use of ADN flexibility to increase the Voltage Stability Margin (VSM) of the transmission system. This is achieved with a multi-level optimization involving a centralized OPF problem at the Energy Management System (EMS) of the TSO and distributed optimization problems at the ADN level similarly to ~\cite{unpublished_PSCC22}. In the problem formulation proposed in ~\cite{unpublished_PSCC22} the local ADN controllers are receiving PQ injection sensitivity information from the solution of the central VSM problem and are attempting to maximize VSM by modifying their PQ injections with the use of local controllers. One disadvantage of this approach is that each ADN uses its resources regardless of its effectiveness to increase the VSM. In the current paper a different multi-stage optimization process is proposed, where the flexibility of each ADN is locally determined and exported to the central controller, which is then maximizing VSM by determining suitable P, Q injection setpoints to ADN controllers, thus achieving a most efficient control.

The rest of this paper is organised as follows. Section II defines the overall problem structure, and Section III presents the power system modelling and the optimization framework for both the transmission and the distribution network. In Section IV, the approaches for defining the operational flexibility of the ADN are described as well as the linear determination of the region for a 2-bus and a 30-bus ADN respectively. Section V includes the application to the Nordic Test System with a simple 2-bus ADN configuration connected to subtransmission buses and presents the results of the VSM optimization problem. Finally, Section VI presents the optimization results for a simple transmission system to which a detailed transmission feeder is connected, as well as the implementation of the individual control setpoints to achieve the desired solution. The last section forms the conclusions of the paper.

\section{Problem Structure}

\begin{figure}[ht]
\includegraphics[width=3.8in]{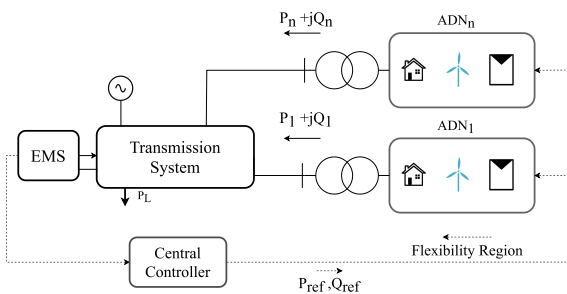}
\caption{Conceptual System Layout}
\label{fig:tso_dso}
\end{figure}

Figure \ref{fig:tso_dso} shows the system layout considered in this paper. It is assumed that each ADN has a dedicated local controller that accepts setpoints and calculates flexibility  boundaries, while a central controller is located at the EMS of the TSO control center. Fig.~1 shows also the information exchange between the centralized and the local ADN controllers.

The Flexibility Region (FR) of each ADN is defined as the area in PQ injection space (at the PCC), for which all operating constraints within the ADN feeder are satisfied. This includes voltage and current constraints in nodes and branches. The boundary of the flexibility region is determined in this paper as the solution of an optimization problem, the objective of which is the maximum deviation from an initial operating point in PQ injection space. This problem is solved by each local ADN controller separately at periodic intervals.

The Central Controller at the TSO EMS level solves the VSM for each contingency and determines the most critical ones. It can then determine P,Q injection setpoints to maximize the VSM for the most critical contingency(ies).
It is noted that the margin maximization may refer either to Voltage Stability of the current operating point, or to post contingency security margin, i.e. the maximum permissible loading after a contingency happens. In the latter case the control determined, i.e. the modification of PQ injections is applied only after the contingency happens~\cite{unpublished_PSCC22}.

Two alternative approaches are proposed in this paper for determining the FR of each ADN. In the first method, the points of multiple binding operational constraints in the distribution feeder are identified and are used to define the boundary of the FR. In the second approach, a radial scan is performed, in order to maximize the distance from the initial operating point in multiple directions subject to all operational constraints, similarly to ~\cite{8917723}. The above two methods are tested in simple and more complex ADN configurations.

The boundary of the FR of each ADN determined periodically by its local controller is exported to the TSO control center in the form of linear constraints on P and Q injection variations around the current operating point. These constraints are then added to the central OPF problem
solved by the central controller, in order to maximize the voltage stability margin by varying the ADN P and Q injections within the specified limits. 

A third and final problem is the activation of ADN controls to achieve the desired setpoint determined by the central OPF problem. In this paper this is formulated as a local optimization problem which tries to minimize the Euclidean distance in PQ space of the actual power injection at the PCC from the desired setpoint, by varying available ADN feeder controls which include IBG and Load Tap Changer (LTC) transformer secondary voltage setpoints, always subject to operational constraints. In the implementation stage the modified LTC setpoint is activated first, followed by a slow ramp modifying IBG voltage and reactive power as in  \cite{MANDOULIDIS2021107438}.

All optimization problems are solved with sequential quadratic programming (SQP) using the "fmincon" function of MATLAB Optimization Toolbox~\cite{MATLAB:2018}.

\section{Centralized OPF and System Modelling}
This section describes the mathematical models used. 
These include the central optimization problem for the VSM estimation and maximization, an approximate approach to include synchronous generator reactive power constraints, the ADN model, and its controls that determine the flexibility region. 

\subsection{Optimization Formulation for Voltage Stability Margin }
Voltage stability margin can be defined as the solution of an optimization problem, where the system load is increased considering a specific stress direction to its maximum, subject to the long term equilibrium conditions~\cite{van2007voltage}. This optimization problem formulation has been presented also in~\cite{unpublished_PSCC22} and is detailed below. 

Let $\mathcal N$ be the set of buses and $G$ the subset of generators. Starting from a given initial operating point, a stress direction is assumed, along which load and generation is increasing with one degree of freedom according to:
\begin{align}
    P_k = P_{0k}+\lambda d_{pk}, k \in \mathcal N \label{(1)}\\
    Q_k = Q_{0k}+\lambda d_{qk}, k \in \mathcal N \label{(2)}
\end{align}
where $\boldsymbol d_p$ and $\boldsymbol d_q$ are the vectors defining the active and reactive power stress direction, and $\lambda$ is a scalar corresponding to the level of stress. 

VSM is obtained by solving the optimization problem for maximizing total load increase:
\begin{equation}
\Delta P = \lambda \sum_{k \in \mathcal N} d_{pk}
\label{DP}
\end{equation}
subject to long-term equilibrium constraints. Since \eqref{DP} has no unconstrained maximum, the maximization corresponds to the point where the long-term equilibrium point disappears, i.e. where the  system undergoes a static bifurcation. Thus, $\Delta P_{max}$, corresponds to the voltage stability margin along the specified stress direction.

The stress direction includes also the distribution of the shift in load among the generators in the system. This is done using a distributed slack approach, so that all generators participate as described below:
\begin{equation}
       P_{gk}=P_{gk0}+\Delta P_{gk}, k\in G
\end{equation}
where the generator adjustments $\Delta P_{gk}$ are determined using participation factors ${w_k}$ normalized to sum up to 1:
\begin{equation}
\sum_{k \in G} w_k = 1
\end{equation}
Total generation must equal total load plus total losses. Thus, 
the loss increase during the stress $\Delta L$
is added to the total load increase to obtain 
generation shifts as: 
\begin{equation}
\Delta P_{gk} = w_k(\Delta L +  \Delta P)
\end{equation}

In the above formulation $\Delta L$ is an extra (slack) variable that in power flow equations is matched-up by setting to zero the imaginary part of the slack bus complex voltage.

The OPF formulation is:
\begin{eqnarray}
\rm{VSM} =\max_{\boldsymbol {e,f,V},\Delta L,\lambda} \Delta P 
\label {(3)}\\
{\rm subject~to:}\nonumber \\
\boldsymbol{g} (e,f,V,\Delta L,\lambda) =0  
\label{g} \\
P_{gk}^{min} \leq P_{gk} \leq P_{gk}^{max}, \forall k \in G  \label{(6)}\\
Q_{gk} \leq q_{ak}(P_{gk},V_k),  \forall k \in G \label{(7)}\\
Q_{gk} \leq q_{rk}(P_{gk},V_k),  \forall k \in G \label{(8)}\\
V_{gk} \leq V_{gk}^{ref}, \forall k \in G
\label{(9)}
\end{eqnarray}
where \eqref{g} is the set of long-term equilibrium equations assuming load restoration to constant power (through e.g. LTC restoration of distribution voltage). Thus, $\boldsymbol{g}$
correspond to the (quadratic) power balance equations in rectangular coordinates $\boldsymbol e, f$.
$P_{gk}$ and $Q_{gk}$ are the active and reactive power generation, $V_{gk}^{ref}$ is the voltage setpoint and $V_{gk}$ is the voltage magnitude of generator $k$:
\begin{equation}
{V_{gk}}^2 = {e_k}^2 + {f_k}^2 
\end{equation}
Active power generation constraints are defined in \eqref{(6)} and are checked at the solution. If the constraints are violated, the limited generators are replaced by constant P injections and the parameters of vector $\boldsymbol{w}$ are updated so that the problem is solved again. Constraints \eqref{(7)},\eqref{(8)} represent the reactive power generation limits, where $q_{ak}(P_{gk},V_k)$ corresponds to the armature current limit and $q_{rk}(P_{gk},V_k)$ to the rotor (field) current limit. Both $q_{ak}$ and $q_{rk}$ are functions of active power $P_{gk}$ and voltage magnitude $V_{gk}$ and are defined in the following subsection. For each generator either \eqref{(9)} or one of \eqref{(7)},\eqref{(8)} must hold as equality. However, there is no need to add one more complementarity constraint, since the load maximization ensures that at the solution the free variable of generator voltage will be at its highest possible value. Again this is checked at the solution.

It is noted that the network equations $\boldsymbol g$ terminate at the primary side of HV/MV transformers feeding ADNs, which are considered as the Points of Common Coupling (PCC) for connected ADNs. Load restoration to constant power is taking place by LTCs not explicitely modelled in $\boldsymbol g$.

Finally, as stated before, the above equations may refer to the current operating state of the system, or to a post contingency condition.

The above optimization problem \eqref{(3)}-\eqref{(8)} can be further modified to maximize VSM by including as control variables the change in P and Q consumption of controlled ADNs. These new variables $\Delta P_j$ and $\Delta Q_j$, are subject to the linear constraints in PQ space defining the flexibility region of each ADN. 

Let $\mathcal{A} \subseteq N$, the subset of nodes where ADNs are connected. Then, active and reactive power consumption is represented as:
\begin{eqnarray}
P_j &=& P_{0j} + \lambda d_{pj} + \Delta P_j, \quad \forall j \in \mathcal{A} \label{(15)}\\
Q_j &=& Q_{0j} + \lambda d_{qj} + \Delta Q_j, \quad \forall j \in \mathcal{A} \label{(16)}
\end{eqnarray}

Each ADN$_j$ 
is subject to a set of $N_j$ linear equations defining  the FR. The number of linear constraints corresponds to the number of straight lines used to approximate the FR as a polygon.
\begin{equation}
\alpha_i \Delta P_j + \beta_i \Delta Q_j +1 \geq 0, 
\quad  i = 1,..,N_j, \quad \forall j \in \mathcal{A}
\label{(17)}
\end{equation}
Coefficients $\alpha_i$ and $\beta_i$ correspond to the line segment between two consecutive vertices of the flexibility polygon and will be defined in Section IV.

The set of equations \eqref{(17)} is added to the constraints of the OPF problem formulation \eqref{g}-\eqref{(8)}, while equations \eqref{(15)}-\eqref{(16)} substitute \eqref{(1)},\eqref{(2)} for nodes connected to ADNs.

The objective function \eqref{DP} does not change, however the optimization now includes the consumption changes as free variables, thus obtaining the maximum VSM:
\begin{equation}
\label{vsm}
\rm{VSM}_{max}=\max_{\boldsymbol {e,f,V, \Delta P, \Delta Q},\Delta L,\lambda} \Delta P 
\end{equation}

\subsection{Synchronous Generator Constraints}
Since the voltage stability margin is heavily affected by the reactive capabilities of synchronous generators, it is crucial to properly represent in the constraints of the optimization problem both armature and field current limits for every synchronous machine $k \in G$. For simplicity of the presentation the subscript $k$ is omitted in the following.
\subsubsection{Armature Current Limits}
If $I_{N}$  is the maximum current limit, then from the definition of apparent power:
\begin{equation}
 I_N=S_{N} / V_N
\end{equation}
So, the inequality constraint \eqref{(7)} is rewritten as:
\begin{equation}
Q_{g} \leq q_a(P_g,V_g) = \sqrt{(V_g I_N)^2-{P_{g}^2}}\label{armature}
\end{equation}

\subsubsection{Field Current Limits - Saturated Machine}
Again the reactive power limit, taking into account the field current limit $E_{lim}$ and magnetic saturation, is a function of generated active power and voltage. Saturation is expressed with the use of saturation factor $K<1$ given by~\cite{van2007voltage}:
\begin{equation}
K = \frac{1}{1+mV_{l}^n}
\end{equation}
where $m,n$ are saturation coefficients and the air-gap flux $V_{l}$ is a function of generator voltage and loading:
\begin{equation}
\hat{V}_l = {V_g} + j X_l \frac{P_g - jQ_g}{V_g} 
\end{equation}
where $X_l$ is the leakage reactance. Note that from the above complex equation only the magnitude of $\hat{V}_l$ needs to be calculated.

Saturation factor $K$ is affecting both saturated emf and saturated d-axis reactance. Thus at the rotor current limit:
\begin{eqnarray}
   E_{qs} & = & KE_{lim}\\
   X_{ds} & = & X_l + KX_{ad}
\end{eqnarray}

\begin{figure}[ht]
\centering
\includegraphics[width=2.8in]{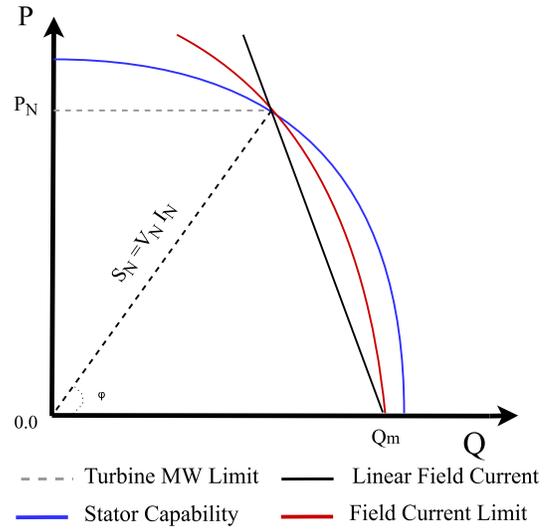}
\caption{Synchronous Round Rotor Machine Capability Curve}
\label{fig:capability}
\end{figure}

In Fig. \ref{fig:capability}, the generator capability curve is shown for nominal voltage $V_N$. The rotor current limit is approximated in this work with a linear approximation 
joining the stator current limit for $P=P_N$ with the point of maximum reactive power generation $Q_m$ for $P=0$ considering the level of saturation of the machine. Thus $Q_m$ is defined as:
\begin{equation}
Q_m = \frac{V_g}{X_{ds}}(E_{qs}-V_g)
\end{equation}
and the linear approximation of the field current limit
in \eqref{(9)} is given by:
\begin{equation}
q_r(P_g,V_g) = Q_m-\gamma(V_g)P_g
\end{equation}
with the slope $\gamma$ given as:
\begin{equation}
\gamma = \frac{Q_m-\sqrt{(V_g I_N)^2-P_N^2}}{P_N}
\end{equation}

\subsection{Active Distribution Networks}
Each ADN$_j, j\in \mathcal A$, is considered separately in this subsection following the configuration of Fig. \ref{fig:tso_dso}. The ADN is absorbing power $P_j, Q_j$ from the primary side of a distribution transformer, while the voltage of the secondary transformer bus is regulated by the transformer LTC. The ADN includes voltage sensitive loads, IBGs, and possibly storage. Let $\mathcal N_{A}$ be the set of the considered ADN buses and $G_A$ the set of buses with IBGs. It is further assumed that BESS can be installed at the same bus as the IBGs, so that the active generation is in general dispatchable.   
The flexibility region of the ADN is defined below.

\subsubsection{Flexibility Region Border Determination}
\begin{figure}[htb]
\centering
\includegraphics[width=3in]{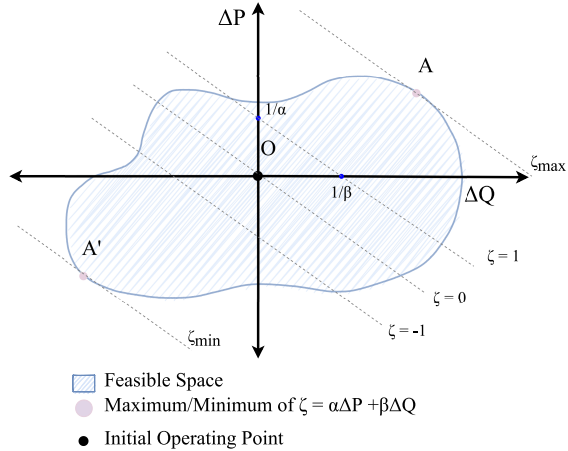}
\caption{Typical FR of an ADN}
\label{fig:feasibility_region}
\end{figure}

The flexibility region border consists of points in PQ space, where the change in power consumption $\Delta P_j, \Delta Q_j$ defined in \eqref{(15)}, \eqref{(16)} meets an operating constraint limit. Thus each point on the FR boundary is a solution of an optimization problem, where a function $\zeta (\Delta P_j, \Delta Q_j)$ is maximized subject to the power balance and ADN operational constraints:
\begin{eqnarray} 
\max_{e,f} \quad \zeta(\Delta P_j,\Delta Q_j)  \label{max2}\\
\boldsymbol{g_A} (e,f) =0 \label{g_A}\\
V_k^{min} \leq V_{k} \leq V_{k}^{max}, \forall k \in \mathcal N_{A}\label{vk} \\
P_{gk}^2+Q_{gk}^2 \leq (V_{k} I_{Nk})^2 ,   \forall k \in G_{A}\label{current_limit} \\
P_{gk}^{min}\leq P_{gk} \leq P_{gk}^{max},  \forall k \in G_{A} \label{bess}\\
V_l^2 = e_k^2 + f_k^2, \forall k \in \mathcal N_{A}
\end{eqnarray}
where, $\boldsymbol{g_A}$ is the set of power balance equations, and \eqref{vk} represent the voltage constraints for every bus of the ADN. Branch flow limits are neglected in this formulation for simplicity, but they can be included in the formulation, if considered necessary. Function $\zeta$ has no unconstrained maximum or minimum points, so the optimal points are determined by the active operational constraints. For example, problems solution for $\zeta=\pm (\alpha \Delta P+\beta \Delta Q)$ ensures a point on the boundary of the feasible solution set as presented in Fig.~\ref{fig:feasibility_region}.

IBG current limits are included in \eqref{current_limit}  similarly to \eqref{armature}.
It is assumed that BESS are connected to the same converter as the distributed generator and for the time duration that will be needed in case of an emergency, the storage combined with the distributed generation can provide active power inside the full range of \eqref{bess}. In the absence of BESS, upper and lower active power limits coincide with the input of the DG source.

\subsubsection{LTC Representation}
The LTC is not explicitly represented and the effect of its deadband is neglected in the OPF formulation that defines the FR boundaries. The regulated voltage $V_d$ of the secondary side of the HV/MV transformer is one of the decision variables. Flows $P_j$,$Q_j$ are calculated at the primary (transmission bus) side to include reactive power losses on the leakage reactance of the transformer $X_t$ (transformer is assumed lossless). For the decision variable $V_d$, equation \eqref{vk} defines its limits together with those of other ADN buses. This assumption is plausible as long as the LTC tap range is not exhausted, so that it can regulate the distribution side voltage close to its setpoint.

\subsubsection{Load Representation}
The loads in the ADN are modelled as voltage sensitive with exponents $a$ and $b$ for active and reactive power. In the examples of this paper active load is constant current $a=1$ and reactive load constant admittance $b=2$.
\begin{eqnarray}
P_{k} &=& P_{0k} (\frac{V_{k}}{V_{0k}}) ^a,   \forall k \in \mathcal N_{A}\label{Pk}\\
Q_{k} &=& Q_{0k} (\frac{V_{k}}{V_{0k}}) ^b,   \forall k \in \mathcal N_{A}\label{Qk}
\end{eqnarray}
Voltage dependency of loads, as presented in \eqref{Pk} and \eqref{Qk} offers the opportunity of indirect load reduction through voltage control. If specific loads are considered dispatchable, meaning that they can actively control their demand upon a request from the control centre, these will be treated as controlled injections $P_g$ on top of their consumption, similar to DG/BESS.

\section{Flexibility Region Estimation}
\begin{figure}[ht]
\centering
\includegraphics[width=3in]{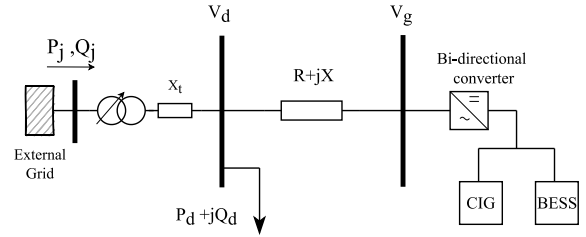}
\caption{One line diagram of 2 bus ADN}
\label{fig:2bus}
\end{figure}

In this section two different approaches are used to estimate the operational flexibility region of two different ADN configurations and then extract equations~\eqref{(17)}. First, the approach of multiple binding constraints is presented and applied on the 2 bus ADN of Fig.~\ref{fig:2bus}. Then, the radial scan approach is examined and used for both the 2 bus ADN and a more complex 30 bus ADN with multiple loads and CIGs.

\subsection{Multiple Binding Constraints}
The first approach developed in order to determine the FR of simple ADN configurations is based on identifying the points on the PQ plane where multiple constraints of the problem \eqref{max2}-\eqref{bess} are simultaneously met, and then linearly connecting them in order to form the FR boundary.  


Points of multiple binding constraints, or corner points (CP) are also examined in~\cite{4282079} in order to identify switching loadability limits, also known as limit induced bifurcations. 
In this section, the corner points are considered as vertices of a polygon in PQ space that bounds the FR.


The ADN configuration of Fig.~\ref{fig:2bus}, also used in~\cite{MANDOULIDIS2021107438}, consists of an aggregate voltage depended load placed on the secondary bus of the HV/MV transformer and a IBG. In this case no BESS is considered and the active power generation $P_g$ of IBGs is constant. So, the decision variables are the two bus voltages $V_d$ of the distribution transformer secondary, and $V_g$ of the IBG, which are both subject to \eqref{vk} for minimum and maximum voltage. Another constraint to be considered is  \eqref{current_limit} for which both a $Q_{max}$ and a $Q_{min}$ limit can be met as the limit depends on $Q^2$. It is noted that if any two of the three variables $V_d, V_g, Q_g$ are determined, the power consumption $P_j, Q_j$ is determined, and thus the change of consumption $\Delta P_j, \Delta Q_j$ which determine the flexibility of the ADN.

With constant $P_g$, active power flexibility depends on $V_d$, thus the limit values of voltage determine the $\Delta P_j$ extreme values. Similarly, $\Delta Q_j$ extremes are linked with $V_g$ or $Q_g$
limits. Each of the two extreme values of $V_d$ can be combined with maximum or minimum of either $V_g$ or $Q_g$, whichever will occur first. This gives $2x2=4$ corner points. Two further corner points correspond to 
$V_g$ and $Q_g$ being simultaneously at maximum or minimum. Thus 6 corner points in general can be determined in the two-bus ADN feeder. 

\begin{figure}[htb]
\centering
\includegraphics[width=3.1in]{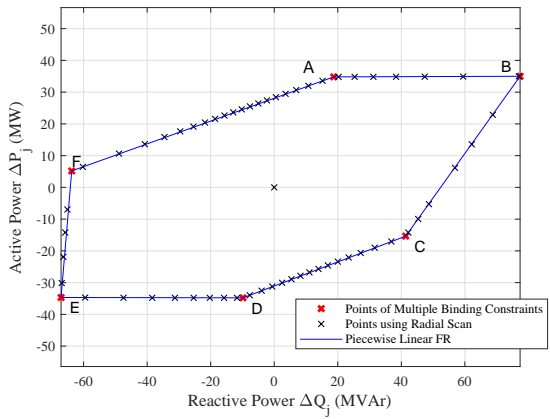}
\caption{Flexibility Region of two bus ADN using the Multiple Binding Constraints Approach}
\label{fig:FR2bus}
\end{figure}

For the feeder data of Table~1 the above six corner points are summarized in Table \ref{table:I} and are shown in PQ space in Fig. \ref{fig:FR2bus}. 

\begin{table}[htb]
\renewcommand{\arraystretch}{1.3}
\caption{Points of Multiple Binding Constraints on the two bus ADN}
\label{table:I}
\centering
\begin{tabular}{|c|c|}
\hline
Point & Set of Binding Constraints \\
\hline \hline
A & $V_d^{max} , V_g^{max}$ \\
\hline
B & $V_d^{max} , -Q_g^{max}$ \\
\hline
C & $V_g^{min} , -Q_g^{max}$ \\
\hline
D & $V_d^{min} , V_g^{min}$ \\
\hline
E & $V_d^{min} , Q_g^{max}$ \\
\hline
F & $V_g^{max} , Q_g^{max}$ \\
\hline
\end{tabular}
\end{table}

The corner points in Fig.~\ref{fig:FR2bus} are joined with linear segments so as to form a polygon. Each line segment represents a linear constraint according to \eqref{(17)}, where the coefficients of each line are:
\begin{equation}
\alpha_i = \frac{\Delta P_{i+1} - \Delta P_i}{\Delta P_i \Delta Q_{i+1}-\Delta P_{i+1} \Delta Q_i}, \quad i = A,..,F
\end{equation}
\begin{equation}
\beta_i = \frac{\Delta Q_{i} - \Delta Q_{i+1}}{\Delta P_i \Delta Q_{i+1}-\Delta P_{i+1} \Delta Q_i}, \quad i = A,..,F
\end{equation}

The figure shown in Fig.~\ref{fig:FR2bus} corresponds to one of the feeders replaced in \cite{MANDOULIDIS2021107438} in the Nordic Test System and the data of the feeder at the initial operating point are presented in Table~\ref{table:II} .
\begin{table}[!ht]
    \caption{Two Bus ADN Data}
    \label{table:II}
    \centering
    \begin{tabular}{|c|c|}
    \hline
        $P_{j0} (MW)$ & 600  \\ \hline
        $Q_{j0} (Mvar)$ & 180  \\ \hline
        $P_d (MW)$ & 696.784  \\ \hline
        $Q_d (Mvar)$ & 142.471  \\ \hline
        $P_g (MW)$ & 97.2  \\ \hline
        $Q_g (Mvar)$ & 0  \\ \hline
        $R (pu)$ & 0.004413333  \\ \hline
        $X (pu)$ & 0.0608  \\ \hline
        $X_t (pu)$ & 0.0015  \\ \hline
    \end{tabular}
\end{table}

It is noted that depending on the two-bus feeder data it is possible for two corner points to be identical or infeasible. In these cases, only the feasible points are used to estimate the FR.

\subsection{Flexibility Region Using Radial Scan}
Identifying the corner points in longer and more complex networks may not be so simple, since  equations and decision variables increase. As a result, a second approach is developed and examined on the two bus feeder and a more realistic feeder of Fig. \ref{fig:30bus}.
This approach estimates the FR by maximizing the distance from an initial operating point along multiple search directions ensuring that the points found in the PQ plane will be on the boundary of the feasible region, since operational constraints are satisfied. For each search direction, two points are calculated, one for maximizing and one for minimizing the change in the active power flow at the PCC of each ADN, similar to the method proposed in~\cite{8917723}. 

The approach is developed as follows:
\begin{enumerate}
  \item The ADN is initialized based on the current operating point where $\Delta P_{j0} = 0, \Delta Q_{j0} = 0$.
  \item Depending on the desired granularity, a small step in the angle $\theta$ between the active and reactive power is chosen, where the tangent is:
  \begin{equation}
    z = tan(\theta),\quad 0 \leq \theta \leq 180
  \end{equation}
  In this work the step in the angle is chosen as
  \begin{equation}
   \Delta \theta = 3^o
  \end{equation}
  and the application on the test systems proves that it provides the needed granularity to define the FR appropriately.
  \item For every step, an optimization problem is solved twice, following the modelling described in Section III.C and equations \eqref{max2}-\eqref{bess} and the addition of the equality constraint \eqref{tangent}:
  \begin{equation}
    \Delta Q_{j} =  z \Delta P_{j} \label{tangent}
  \end{equation}
    \item The solution of the OPF problem at each step provides two points by maximizing and minimizing   \begin{equation}
         \zeta =\pm (\alpha \Delta P+\beta \Delta Q)
     \end{equation}
     where, $\alpha =1$, $\beta = 0$
\end{enumerate}

\begin{figure}[ht]
\centering
\includegraphics[width=3in]{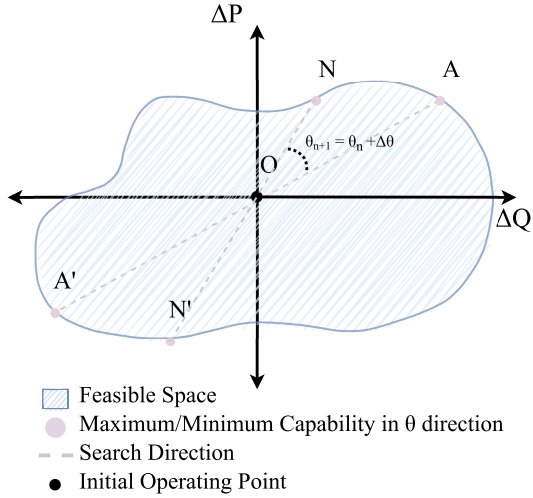}
\caption{Feasiblity Region using the Radial Scan}
\label{fig:FR_scan}
\end{figure}

\begin{figure}[ht]
\centering
\includegraphics[width=3.2in]{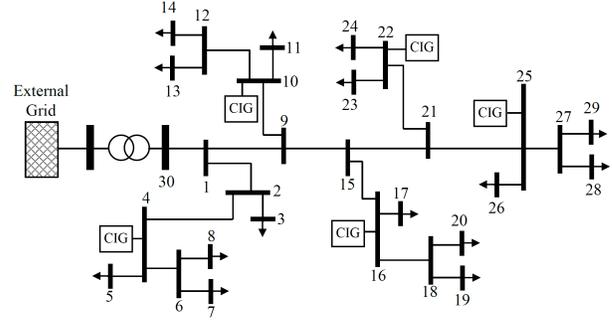}
\caption{One line diagram of 30 bus ADN}
\label{fig:30bus}
\end{figure}

The constraints of the optimization problem ensure that each solution will be feasible and in the boundary of the flexibility region. In Fig.~\ref{fig:FR_scan}, points A,N and A',N' represent points on the boundary of the FR for different angles $\theta$ for positive and negative values of function $\zeta$.

Depending on the number of steps of the change in the search direction, multiple points are found. Using all points would result in numerous sets of constraints for each ADN according to \eqref{(17)}. As a result, an algorithm is created in order to keep only the points that will best define the FR and ensure that it will be convex. The FR is again created by linearly connecting the points.

$\boldsymbol{Step} \quad \boldsymbol{1}$ : Initially, three points are selected. It is crucial that the point $[\Delta Q_j^0, \Delta P_j^0]$ is inside the triangle that is formed by the three points. The points used are $\Delta Q_j^{min},\Delta Q_j^{max},\Delta P_j^{min}$. Selection of 3 points ensures that the resulting FR will be convex, as the points added will be external of the already convex triangle.

$\boldsymbol{Step} \quad \boldsymbol{2}$ : The algorithm only adds points that are outside the existing area. So, for each line segment between the initial points, all the points where \eqref{(17)} is negative are found.

$\boldsymbol{Step} \quad \boldsymbol{3}$ : Between the points calculated in the previous step, the one with the greatest distance from the respective line segment is kept, where the distance of all points from line segment $i$ is $\boldsymbol{d}$:

\begin{equation}
{d}_i = \frac{|\alpha_i \Delta P_{ji} + \beta_i \Delta Q_{ji} +1|}{\alpha_i^2 + \beta_i^2}
\end{equation}

$\boldsymbol{Step} \quad \boldsymbol{4}$ : A single point is added to the three initial points and the procedure returns to Step 2. In this work six points are considered enough to approximate the FR.

The proposed approach is applied on both the 2 bus ADN of Fig.~\ref{fig:2bus} and the 30 bus ADN of Fig.~\ref{fig:30bus}.
\subsubsection{2-bus ADN}
In Fig.~\ref{fig:FR2bus} the FR of the 2 bus ADN is presented using both methods, meaning that for simple configurations where it is easy to identify the points of multiple binding constraints, both approaches present a good approximation of the FR.
\subsubsection{30-bus ADN}
The 30-bus, 20kV distribution system~\cite{9494823} is connected to the transmission system through a 220/20kV, 20 MVA, transformer. It is slightly modified by replacing synchronous machines with IBGs and using voltage depended loads in all load buses.
\begin{figure}[ht]
\centering
\includegraphics[width=3.1in]{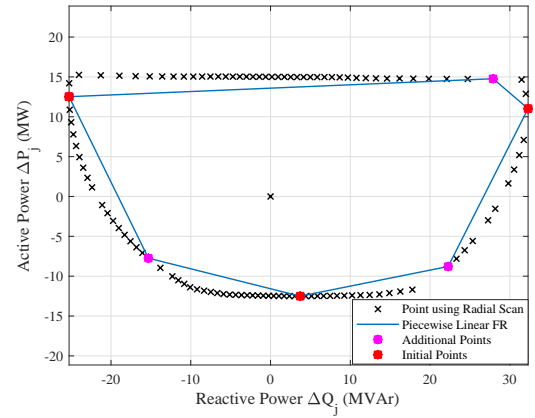}
\caption{Piecewise Linear FR of 30 bus ADN including storage}
\label{fig:linear1}
\end{figure}
\begin{figure}[ht]
\centering
\includegraphics[width=3.1in]{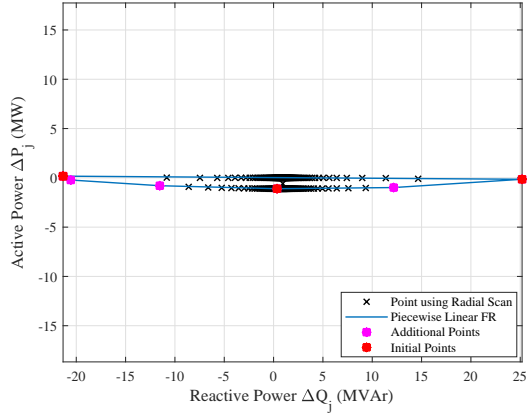}
\caption{Piecewise Linear FR of 30 bus ADN without storage}
\label{fig:linear2}
\end{figure}
In Figures \ref{fig:linear1} and \ref{fig:linear2}, the piecewise linear FRs for the 30 bus ADN of Fig. \ref{fig:30bus} are presented. In the first case, storage is considered while in the second there is no storage, so the change in active power only relies in the indirect change due to the voltage dependence of the loads. Also, reactive power is influenced by the active power generation of CIGs due to \eqref{current_limit}.
For the 30 bus ADN, each directional OPF problem is solved in about 20 seconds in a laptop with GPU Core i7, 1.8 GHz and 8 GB RAM. This speed allows to compute the whole FR periodically (e.g. every hour) and can be improved using more efficient techniques.
\section{Application on the Nordic Test System}
\begin{figure}[ht]
\centering
\includegraphics[width=3.1in]{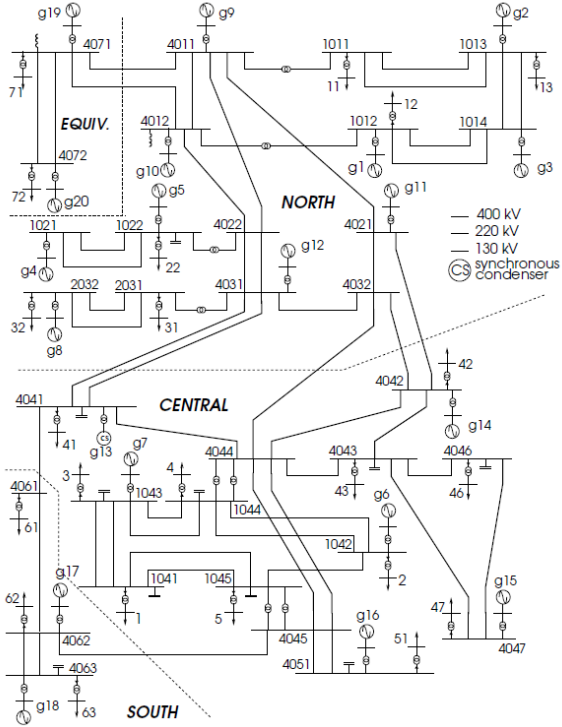}
\caption{One line diagram of the IEEE Nordic Test System}
\label{fig:nordic}
\end{figure}
\subsection{System Description}
The central optimization problem, posed in Section III.A, is applied to the IEEE Nordic Test System~\cite{9018172} of Fig.~\ref{fig:nordic}, which is modified similarly to~\cite{MANDOULIDIS2021107438}, but with only the five feeders connected to the subtransmition HV buses of the Central Area replaced by ADNs. These ADNs are represented with the two-bus equivalent of  Fig.~\ref{fig:2bus} with the data and initial conditions given in~\cite{MANDOULIDIS2021107438} and the initial $P,Q$ consumption shown in Table~\ref{table:III}. 

\begin{table}[htb]
\caption{Initial Consumption by ADNs}
\label{table:III}
\centering
\begin{tabular}{c c c }
{Bus} & {$P_j$ (MW)} & {$Q_j$ (MVar)}  \\ 
\hline
1041         & 600              & 180                 \\
1042         & 330              & 90                  \\ 
1043         & 260              & 100                 \\
1044         & 840              & 300                 \\
1045         & 720              & 230                 \\
\end{tabular}
\end{table}
The system is prone to voltage instability due to high power transfers from the north to the central area.
In~\cite{unpublished_PSCC22}, several critical contingencies that can result in voltage instability have been identified. In this paper, only the most critical contingency will be examined, which is the loss of transmission line 4032-4044. 

The direction of stress chosen for the VSM determination, is a uniform load increase in the central area only, with all generators participating resulting in an increase of power transfer to the Central Area which is generation deficient.
Thus $d_{pk}={P_{0k}}$ and $d_{qk}={Q_{0k}}$ while all other elements of $\boldsymbol{d_p}$ and $\boldsymbol{d_q}$ are set to zero.
. Thus $d_{pk}={P_{0k}}$ and $d_{qk}={Q_{0k}}$ while all other elements of $\boldsymbol{d_p}$ and $\boldsymbol{d_q}$ are set to zero.

\subsection{ADN Flexibility Regions}
Following the approach of multiple bindings constraints, presented in Section IV, the flexibility region is calculated for the five ADNs of Table~\ref{table:III}. 
For each ADN the flexibility region is determined through the linear constraints \eqref{(17)}, and the relevant coefficients $\alpha$ and $\beta$ are presented in Table~\ref{table:V}.

\begin{figure}[htb]
\centering
\includegraphics[width=3.1in]{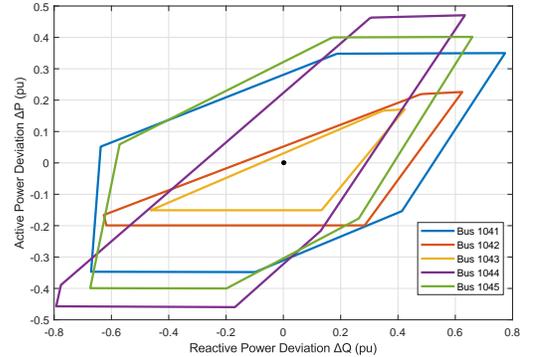}
\caption{Flexibility Regions of 5 ADNs}
\label{fig:5_FR}
\end{figure}

In Fig.~\ref{fig:5_FR}, the FRs of the five ADNs are shown graphically. Depending on the ADN parameters, 
FRs are different for each case.
Note that in some of the cases, some of the points corresponding to multiple binding constraints are infeasible, resulting in fewer linear constraints.

\begingroup
\begin{table}[ht]
\caption {VSM with and without ADN flexibility}
\label{table:IV}
\centering
\begin{tabular}{c c c c}
& Central Area (MW)&  max stress level $\lambda$ & VSM (MW)\\
\hline
Initial Loading  & 6190 & $-$     &  $-$\\
No ADN & 6085 & -0.0169 & -105\\
1 ADN  & 6153 & -0.0058 &  -37\\ 
2 ADN  & 6178 & -0.0020 &  -12\\ 
3 ADN  & 6206 &  0.0027 &   16\\ 
4 ADN  & 6274 &  0.0135 &   84\\ 
5 ADN  & 6329 &  0.0255 &   139\\
\end{tabular}
\end{table}
\endgroup
\begin{table*}[!htbp]
\caption {Feeder coefficients $\alpha$ and $\beta$ that form the linear constraints}
\label{table:V}
\centering
\begin{tabular}{cccccccccc}
\multicolumn{2}{c}{1041} & \multicolumn{2}{c}{1042} & \multicolumn{2}{c}{1043} & \multicolumn{2}{c}{1044} & \multicolumn{2}{c}{1045} \\
\hline
$\alpha$           & $\beta$          & $\alpha$           & $\beta$           & $\alpha$           & $\beta$           & $\alpha$            & $\beta$          & $\alpha$            & $\beta$          \\
\hline
0.009  & -2.878 & 0.239  & -5.084  & 0.379  & -6.795  & 0.051  & -2.193 & 0.008  & -2.504 \\
-1.910 & 1.365  & -2.256 & 1.806   & -3.712 & 3.352   & -3.461 & 2.535  & -2.600 & 1.781  \\
-1.221 & 3.219  & 0.000  & 5.022   & 0.000  & 6.631   & -2.518 & 3.102  & -1.586 & 3.287  \\
0.005  & 2.873  & 1.490  & 0.401   & 12.922 & -33.018 & 0.009  & 2.172  & 0.005  & 2.498  \\
1.558  & -0.131 & 6.681  & -19.200 &        &         & 1.468  & -0.358 & 1.710  & -0.383 \\
1.279  & -3.562 &        &         &        &         & 3.535  & -4.483 & 1.428  & -3.112
\end{tabular}
\end{table*}
\begingroup
\begin{table}[ht]
\caption {Active and Reactive power change in each ADN at the VSM solution}
\label{table:VI}
\centering
\begin{tabular}{ccc}
Bus  & $\Delta P$ (MW)      & $\Delta Q$ (MVAr) \\
\hline
1041 & -34.7 & -70.8                  \\
1042 & -19.9 & -61.7                \\
1043 & -15.0 & -46.4                \\
1044 & -45.7 & -79.4                \\
1045 & -39.9 & -67.7                \\
Total& -155.3 & -326.1              \\
Actual $\Delta P_{tot}$& 138 & 
\end{tabular}
\end{table}
\endgroup
\subsection{VSM Results}
The VSM optimization problem \eqref{vsm} is solved for the test system first without ADN flexibility, and then successively by introducing one by one the five ADN flexibility regions. The corresponding maximum VSM and level of stress, along with the total load of the central area, after the most critical contingency are presented in Table~\ref{table:IV}. As expected, without ADN flexibility the system is insecure having a negative post-contingency margin of -105 MW. However, if the three first feeders are assumed to be flexible system security is restored, as the post-contingency VSM becomes positive. Making use of the flexibility of all five ADNs results in a relatively safe margin of 139 MW.

For the last case, where all 5 feeders are flexible, the optimal consumption adjustments $\Delta P_j$ and $\Delta Q_j$ for each ADN$_j$ as provided by the solution of the optimization problem are presented in Table~\ref{table:VI}. In the optimal solution for all five ADNs the load bus voltage is minimized, while either the maximum voltage or the current limit constraint is active for the CIG bus. Thus the optimal solution for each ADN corresponds to either point A, or E of Table~\ref{table:I}. As a result, it is pretty straightforward to determine the optimal setpoints that need to be applied to each feeder to achieve the changes presented in Table~\ref{table:VI}. It is noted that this optimal solution is the same with the intuitive approach employed in \cite{MANDOULIDIS2021107438} and \cite{9210043}, where LTC voltage is minimized and reactive support from CIGs is maximized through a ramp.

\subsection{Simulation of Optimal Solution}
In this subsection, a long-term simulation is performed for the Nordic test system with WPSTAB software developed in NTUA. The simulation includes control actions to achieve the optimal consumption adjustment for the five flexible ADNs.
The considered contingency (loss of transmission line 4032-4044) is applied 30s after the start of the simulation. Since the contingency is identified as insecure (according to Table~\ref{table:V}), immediately after the contingency emedial actions are applied simultaneously to all five ADNs. 
\begin{figure}[ht]
\centering
\includegraphics[width=3.1in]{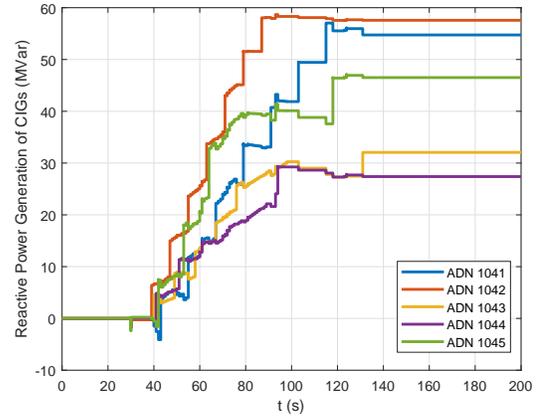}
\caption{Reactive Power Generation of CIGs}
\label{fig:cig_reactive}
\end{figure}
\begin{figure}[ht]
\centering
\includegraphics[width=3.1in]{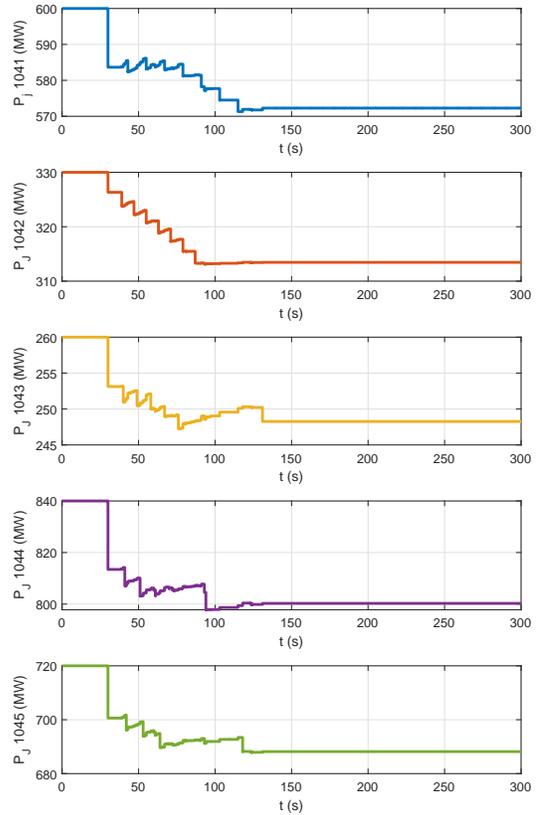}
\caption{Active Power Injections of ADNs in the PCC}
\label{fig:active_power_injection}
\end{figure}
\begin{figure}[ht]
\centering
\includegraphics[width=3.1in]{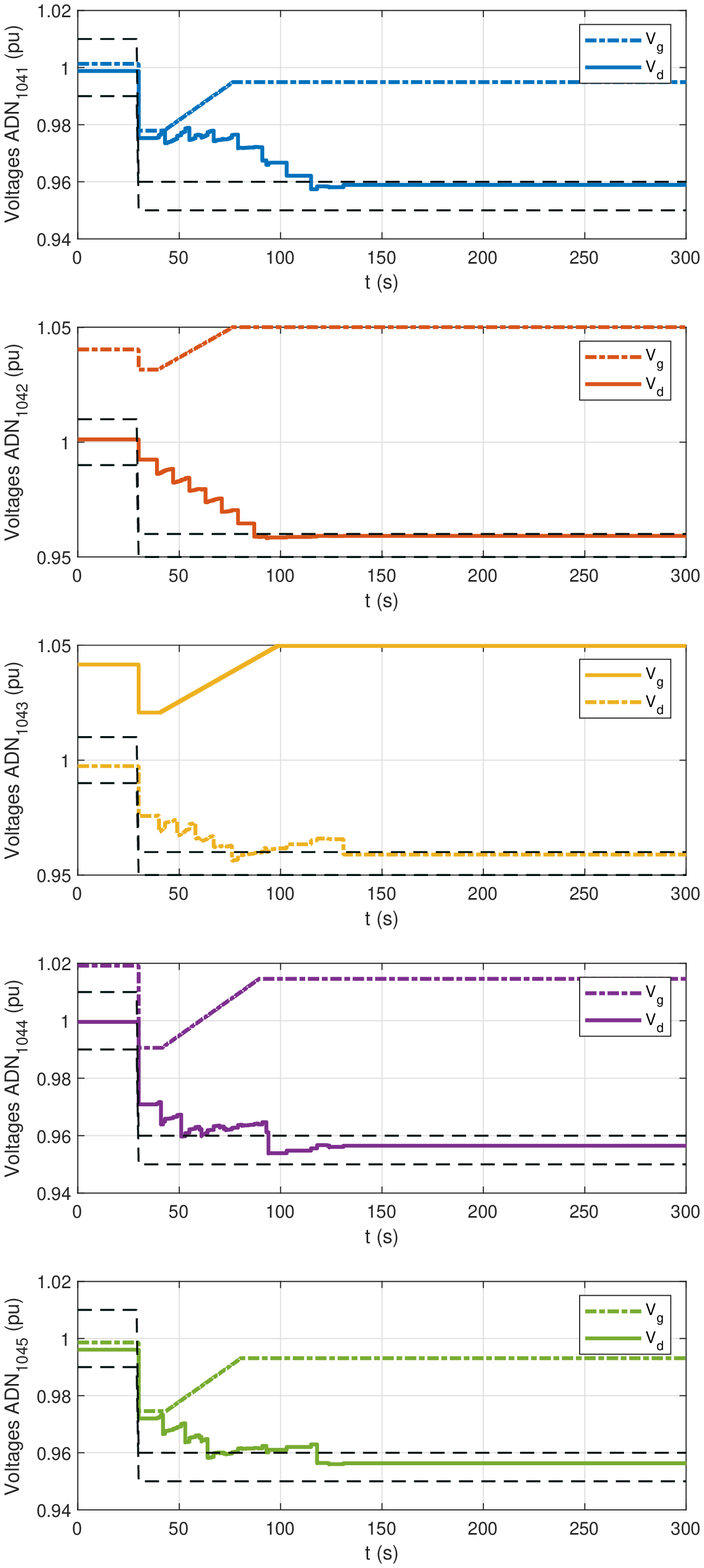}
\caption{Voltages $V_d$ and $V_g$ and LTC deadband}
\label{fig:voltages_adns}
\end{figure}

In each ADN, the voltages calculated in previous subsection in order to achieve the $\Delta P_j$ and $\Delta Q_j$ of Table~\ref{table:VI} are applied as setpoints to the LTC controlled voltage and the voltages of the CIGs. 
More specifically, immediately after the contingency the deadband of the MV bus of the transformer of each ADN is changed. Then, after the first LTC step the voltage setpoint of the corresponding CIG is changed using a voltage ramp of $0.0005 pu/s$ until it achieves the desired value obtained by the optimization. 

The resulting reactive power generation of the CIGs is plotted in Fig.~\ref{fig:cig_reactive}. Due to the voltage dependence of loads~\eqref{Pk},~\eqref{Qk}, the reduction of the MV bus voltage indirectly reduces the active power flow in the PCC, so $\Delta P_j$ for each ADN is presented in Fig.~\ref{fig:active_power_injection}. The voltages $V_d$ of the secondary side of the HV/MV transformer and $V_g$ of the CIG are shown in Fig.~\ref{fig:voltages_adns} for each ADN.  

\begin{figure}[htb]
\centering
\includegraphics[width=3.1in]{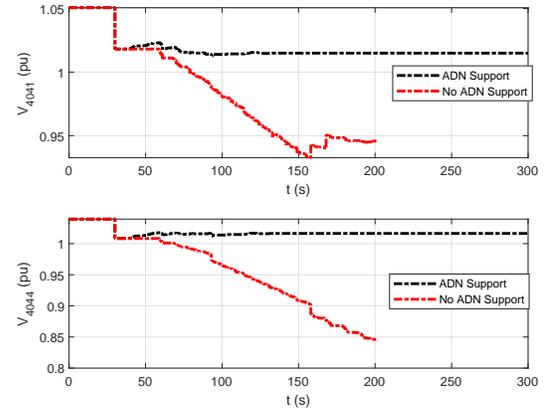}
\caption{Voltages in HV buses 4041, 4044}
\label{fig:hv_voltages}
\end{figure}

Finally, in Fig.~\ref{fig:hv_voltages} the voltages of EHV buses 4041 and 4044 of the central area are presented, with and without the remedial actions implemented on the flexible ADNs, in order to show how the voltage collapse is avoided and a stable operating point is restored.

The amount of indirect load shedding in each feeder $j$ can be calculated using the following equation \cite{MANDOULIDIS2021107438}:
\begin{equation}
\Delta P_j = \frac{P_{kj}^0}{V_{dj}^a}(V_{min,j}^a-V_{fin,j}^a)
\end{equation}
where, $V_{dj}^a$ is the initial load bus voltage,  $V_{fin,j}^a$ is the same voltage at the end of the simulation, and $V_{min,j}^a$ is the lower limit of the initial LTC deadband. The total indirect load reduction amounts to $\Delta P_{tot}=138$~MW,  as seen in the last row of Table~\ref{table:VI}. This is less than the one calculated by the VSM optimization, as the voltage of the load bus is not exactly equal to the specified setpoint due to the deadband.
It is noted that the indirect load reduction is less than the one reported in \cite{MANDOULIDIS2021107438} in the majority of the protection schemes that were used, while keeping both tranmission and distribution voltages inside the operational limits.

\section{Radial Test System with 30-bus ADN}

In this Section the 30-bus feeder of Fig.~\ref{fig:30bus} is considered and the VSM optimization is applied in a simple radial test system, in order to demonstrate how to implement the optimal power adjustment setpoints in more complex feeders.
For this reason, a third OPF problem is formulated, as already discussed in Section II, in order to achieve ADN $P,Q$ consumption as close as possible to the values determined by the central optimization $P_{ref}$,$Q_{ref}$ for the $j^{th}$ ADN:
\begin{eqnarray}
P_{ref}& =& P_j^0 + \Delta P_j\\
Q_{ref}& =& Q_j^0 + \Delta Q_j
\end{eqnarray}
The ADN operational constraints (28)-(31) remain the same, while the objective function to be minimized is the Euclidean distance between the reference point and the actual active and reactive power flows at the PCC:
\begin{equation}
\min_{e,f,V} \sqrt{(P_{ref}-P_j)^2+(Q_{ref}-Q_j)^2}) 
\label{(44)}
\end{equation}

The solution of problem \eqref{(44)} ensures that the ADN will move as close as possible to the reference point inside its feasible region, even if due tot he approximations made in the FR determination and the centralized OPF solution, it may not achieve the exact reference values. 

\begin{figure}[htb]
\centering
\includegraphics[height=2.1in]{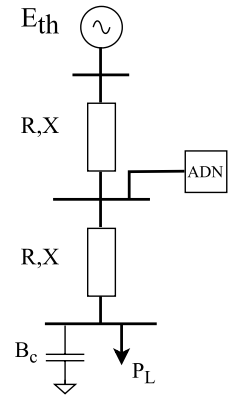}
\caption{Transmission Corridor with ADN connected in the middle}
\label{fig:trans_line}
\end{figure}

The simple radial transmission test system considered is shown in Fig.~\ref{fig:trans_line} with the data provided in Table~\ref{table:VII}. The ADN of Fig.~\ref{fig:30bus} is connected in the middle of the radial transmission line of 
Fig.~\ref{fig:trans_line} which consists of two parts with identical $R,X$.
For this test system, problem~\eqref{vsm} is solved to provide the maximum active power $P_L$ that can be transferred to the remote load bus. The problem is solved first for constant distribution system setpoints, and then again allowing the ADN consumption to vary within the linear FR approximation previously calculated in Fig.~\ref{fig:linear1}. The solution of the optimization provides the maximum power results shown in Table~\ref{table:IX}.
The setpoints for the controllable voltages of LTC and CIGs in order to achieve these reference values are obtained solving the optimization problem \eqref{(44)} and are presented in Table~\ref{table:VIII}.
\begin{table}[ht]
\centering
\caption {Radial Transmission System Data (100MVA base)}
\label{table:VII}
\begin{tabular}{|c|c|c|c|} \hline
$R$ (pu) & $X$ (pu) & $B_c$ (pu) & $E_{th}$ (pu) \\ \hline
0.05   & 0.2  & 2   & 1.05      \\ \hline
\end{tabular}
\end{table}
\begin{table}[ht]
\centering
\caption {VSM Results}
\label{table:IX}
\begin{tabular}{lcc}
          & With ADN & Without ADN \\ \hline
$P^0$ (MW) & -3.24 & -3.24              \\
$Q^0$ (MVAr) & 5.29 & 5.29              \\
$\Delta P$ (MW)   & -6.85    & -           \\
$\Delta Q$ (MVAr) & -20.34   & -          \\
$P_{ref}$ (MW) & -10.09 & -             \\
$Q_{ref}$ (MVAr) & -15.05 & -              \\
VSM (MW)  & 124.59   & 113.34      \\
\end{tabular}
\end{table}

The results show that for this ADN configuration, the VSM is greater if the ADN is able to provide the maximum reactive power from its assets, while respecting \eqref{current_limit}. Initially, CIGs are operating with unity power factor.
\begin{table}[htb]
\centering
\caption {Optimal Setpoints for 30 Bus ADN}
\label{table:VIII}
\begin{tabular}{cccccccc}
\multicolumn{1}{l}{} & \multicolumn{3}{c}{Initial} & \multicolumn{3}{c}{Optimal} \\ \hline
Bus &     $S_{nom}$            & $V_{set}$          & $P_g$     & $Q_g$    & $V_{set}$ & $P_g$ & $Q_g$     \\ \hline
30  &    -           & 1.0000     & -      & -     & 1.0047    & -      & -      \\
4   &    8.8         & 0.9961     & 4.0    & 0     & 1.0383    & 6.43   & 6.49   \\
10  &    6.6         & 0.9906     & 3.3    & 0     & 1.0394    & 4.88   & 4.82   \\
16  &    5.5         & 0.9873     & 2.0    & 0     & 1.0448    & 4.13   & 4.00   \\
22  &    2.75        & 0.9876     & 2.0    & 0     & 1.0489    & 2.08   & 1.99   \\
25  &    5.5         & 0.9861     & 3.0    & 0     & 1.0500    & 4.21   & 3.96  
\end{tabular}
\end{table}

\begin{figure}[htb]
\centering
\includegraphics[height=2.1in]{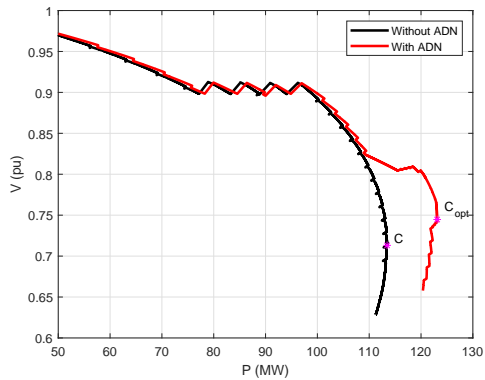}
\caption{PV curve for remote load bus with and without ADN support}
\label{fig:PV}
\end{figure}

The test system is also simulated using long-term simulation of a conductance load ramp at the remote bus past the maximum power transfer limit. The resulting remote load PV curves are presented in Fig.~\ref{fig:PV}, with and without implementing the optimal setpoints already shown in Table~\ref{table:VIII}. The maximum loading points as shown in this figure are $C = 113.4 MW$ and $C_{opt} = 123.2 MW$, which are very close to the VSM solution reported in Table~\ref{table:IX}. While the increase in the VSM is rather small (about 11 MW), compared to the total capacity of the CIGs in the distribution feeder (29.15 MVA) it amounts to about 38 \%, which is a signigicant percentage.

\section{Conclusion}
This paper proposes a framework in order to identify the operational flexibility of Active Distribution Networks and exploit this in order to increase the voltage stability margin of the transmission system. The approach relies on calculating the flexibility region which is translated into the active and reactive power capabilities of each ADN. 

Two different approaches have been presented. The first approach is easier to implement in small feeders with few (e.g. two) controllable voltages. The second approach is general and can work on any ADN regardless of the amount of controllable assets. This is showcased using both a 2-bus and a 30-bus distribution feeder.

The optimization approach using ADN feasibility was applied to the IEEE Nordic Test System extended with simple 2-bus distribution feeders including CIGs, as well as to a simple radial transmission system on which the 30-bus distribution feeder was connected. 

In the VSM problem formulation on the Nordic Test System, five of the feeders were represented using the flexibility regions calculated by assuming multiple binding constraints. The results of the VSM formulation were subsequently validated using time domain long-term simulation, in order to check how the reference values of $P_{ref}$ and $Q_{ref}$ can be implemented via time dependent controls.

The results show that the proposed approach was successful in extending voltage stability margin and avoiding a possible voltage collapse, while simultaneously keeping the operation of the distribution feeder within acceptable operational limits. This results in a non-intrusive framework with significant potential for real system operation.

In this work, optimization problems are solved using a non-linear solver as mentioned in Section II. Nonetheless, convex relaxations of the OPF formulation are considered for future work, in order to increase solution speed and improve convergence.

\FloatBarrier
\bibliography{references.bib}
\bibliographystyle{unsrt}
\end{document}